# Electrochemical synthesis of highly ordered nanowires with a rectangular cross-section using an in-plane nanochannel array

Philip Sergelius[1], Josep M. Montero Moreno[1], Wehid Rahimi[1], Martin Waleczek[1], Robert Zierold[1], Detlef Görlitz[1] and Kornelius Nielsch[1]

psergeli@physnet.uni-hamburg.de
jmontero@physnet.uni-hamburg.de

[1] Institute of Applied Physics, Universität Hamburg, Jungiusstraße 11, 20355, Hamburg, Germany

ABSTRACT

Rapid and reproducible assembly of aligned nanostructures on a wafer-scale is a crucial, yet one of the most challenging tasks in the incorporation of nanowires into integrated circuits. We present the synthesis of a periodic nanochannel template designed for electrochemical growth of perfectly aligned, rectangular nanowires over large areas. The nanowires can be electrically contacted and characterized *in situ* using a pre-patterned multi-point measurement platform. During the measurement the wires remain within a thick oxide matrix providing protection against breaking and oxidation. We use laser interference lithography, reactive ion etching and atomic layer deposition to create cm-long parallel nanochannels with characteristic dimensions as small as 40 nm. In a showcase study pulsed electrodeposition of iron is carried out creating rectangular shaped iron nanowires within the nanochannels. By design of the device, the grown wires are in contact with an integrated electrode system on both ends directly after the deposition. No further processing steps are required for electrical characterization, minimizing the risk of damage and oxidation. The developed nanowire measurement device allows for multi-probe resistance measurements and can easily be adopted for transistor applications. The guided, in-plane growth of electrodeposited nanowire arrays which are tunable in size and density paves the way for the incorporation of nanowires into a large variety of multifunctional devices.

## 1. Introduction

After 20 years of intensive research on one-dimensional nanostructures, we are now on the edge of being able to use nanowires for industry purposes. Several publications have been made exploring their fundamental properties and their broad range of potential applications. In particular, one-dimensional nanostructures have been used for nanowire transistors [1-3], whereas Lavrijsen, Allwood and Cowburn *et al.* have developed logic concepts and models for storage of multiple bits within one nanowire [4-6]. On the field of thermoelectrics enhanced performance due to confinement effects has been theoretically discussed and reported [7,8]. For industry oriented designs it is crucial to be able to accurately position nanostructures into large arrays and naturally there is an everlasting quest for cheap and rapid production, while maintaining a high durability of the fabricated devices [9,10]. Menke *et al.* coined the phrase of lithographically patterned nanowire electrodeposition by presenting an approach that combines surface machining and electrodeposition [11]. Nanowires fabricated using this or similar approaches have already been used for multifunctional sensing devices [12-14]. A large number of other methods for nanowire growth exist, such as vapor-liquid-solid growth [15-17], pulsed laser deposition [18] and the very common template assisted electrodeposition [19-22]. Hundreds of publications which are closely related to the electrodeposition of nanowires are published every year. Apart from a limited number of exceptions in which nanowires are contacted post-growth with two-point contacts within their vertical synthesis template [23-25], all of these techniques have in common that nanowires have to be transferred from their template of growth and placed on a substrate before electrical characterization or device fabrication can be carried out. This involves several processing steps and depending on the growth method, at least one chemical etching step [26] or a method of picking up [27] the wires from the substrate. If the wires need to be aligned, different methods using electric or magnetic fields as well as elaborate nanowire printing systems have been reported [3,9]. However, none of these methods can achieve perfection and all of these methods can damage or alter the nanowires. In any case, photolithography becomes necessary for the connection process. The electrical connection between electronics and nanowires can either be established by time-consuming and often challenging e-beam lithography or by direct lithography [28]. Unless all synthesis steps can be successively conducted without a vacuum break, the sample is exposed to ambient air, humidity or process chemicals which can lead to unwanted oxidation or etching of the nanowires. In the case of bismuth nanowires for example, a strong surface oxide layer is formed upon release from the template which needs to be removed before an ohmic contact between the deposited contact pads and the wire becomes possible [29,30]. In summary, all these steps until its final incorporation into a measurement device pose a significant challenge. The question we seek to answer is how the whole procedure can be simplified.

Herein we present a novel template design that comes with a two-fold advantage: If a limited number of nanowires need to be grown for material characterization, all of the outlined difficulties can be overcome and the characterization is greatly facilitated. Additionally, the approach opens up the possibility to create wafer-scale arrays of parallel nanowires.

Our approach is based on the use of a periodic nanochannel template similar to those frequently used in micro- and nanofluidics [31-34]. However, we do not aim to investigate the fluidic properties, but rather to use these kinds of templates for nanowire fabrication, similar to the works of Pevzner *et al.,* who used VLS-deposition techniques [35]. In contrast to other approaches for in-plane nanowire growth, this has the advantage that the growth direction of the nanowires is perfectly deterministic and confined in all directions but in length, which guarantees a high structural homogeneity. Electrodeposition is known to produce nanowires of excellent quality and since our nanowires have a rectangular cross section, they become particularly interesting from a magnetic point of view [24]. The production of the lithographically defined nanowire arrays is cheap, rapid, requires only standard laboratory equipment and no strongly toxic chemicals. Thus the whole process can easily be scaled to industrial sized device fabrication.

The idea behind the setup is to pattern a pair of flat electrodes onto a substrate prior to the fabrication of an in-plane nanochannel structure on top of these using a sacrificial-layer method [36-38]. This layer is patterned using UV laser interference lithography (LIL) and reactive ion etching (RIE). Subsequently it is covered by atomic layer deposition (ALD). One of the electrodes serves as a cathode in an electrodeposition process, which triggers nanowire growth within the nanochannel system. Once the second electrode is reached during wire deposition, the process is intentionally stopped and by then, electric contact on both ends of the nanowire has been established. The dimensions of the wires are tunable in width, thickness and length by modifying the shape of the nanochannels and the arrangement of the electrodes. Electric measurements can be performed *in situ*, since there is no need of any additional processing steps. During the measurements the nanowires may remain inside the template which protects the wires against oxidation and physical damage. Still, any contact structure for example for multi-probe measurements, thermometers and source-drain layouts can be incorporated into a single system-on-chip device.

## 2. Results and Discussion

The synthesis route of our template is displayed in Figure 1.

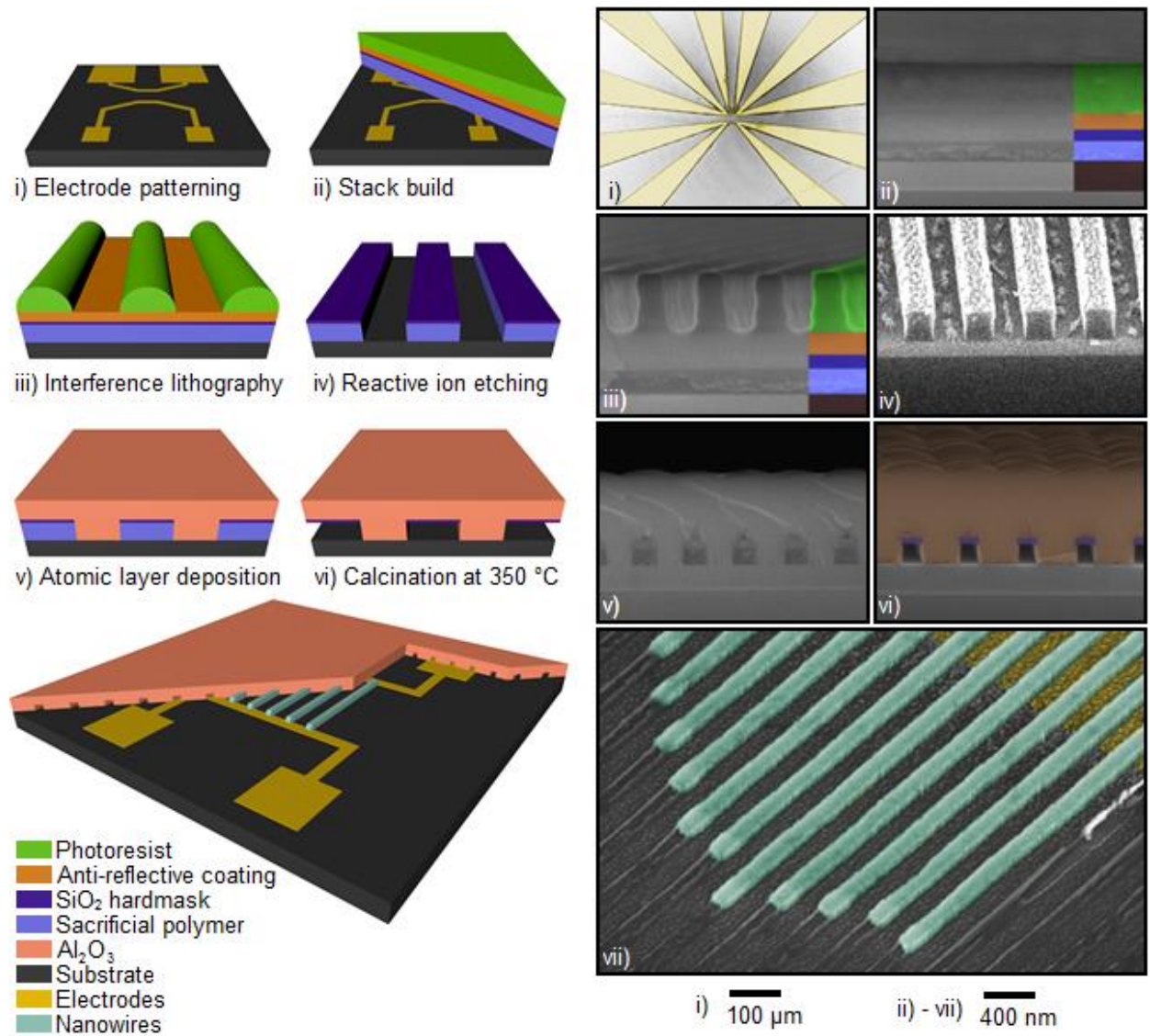

i) 100 μm        ii) - vii) 400 nm

**Figure 1:** Left: Schematic of the synthesis route. Note that all proportions are severely distorted for easier understanding. Right: Colored SEM images of the corresponding processing steps. Image vii) shows short rectangular 80x80 nm² nickel wires which are connected on one side only. The surrounding $Al_2O_3$ template was chemically etched in 0.1 M NaOH.

For our showcase device, an electrode design for 4-probe measurements is drawn in a positive photoresist layer by means of direct laser writing lithography. After sequential sputtering of Cr (adhesion promoter) and a thin Au layer, the photoresist is lifted off with acetone leaving the contact electrode platform on the substrate. Two of the contact pads are designed to function as the working electrode in the later electrodeposition process. The width of the contact pads is made large enough (~1 cm) to ensure easy connectability to the potentiostat (Figure 1.i).

Subsequently a stack consisting of four different layers is prepared in the following order (Figure 1.ii): 1) Spin coating of a sacrificial polymer layer. A suitable polymer is chosen and the spin rate is adjusted to get the desired thickness from 40 nm up to approximately 1 µm. The thickness of this layer determines the height of the nanochannels. Note that any polymer can be used, provided it withstands all further processing steps and can be removed selectively. 2) Sputtering of a thin $SiO_2$ layer acting as a hard mask for pattern transfer. 3) Spin coating of an antireflective coating which is necessary for interference lithography [39]. The spin rate is adjusted to obtain the desired thickness. 4) Spin coating of positive photoresist.

The photoresist is exposed using a laser interference lithography setup based on the Lloyd's-Mirror configuration (Figure 2) [39,40], however an e-beam setup is equally suitable for the line pattern writing. LIL offers great advantages over common illumination techniques with regard to production speed, since wafer-scale areas can be processed in a single exposure. Another upside of LIL is its ability to create significantly smaller feature sizes than any other optical apparatus using the same wavelength, thus greatly reducing the cost for lasers and optics [41]. The interference pattern of a single exposure with one mirror consists of an array of parallel lines with a periodicity given by $P = \frac{\lambda}{2\sin(\theta)}$.

A rotation of the stage with respect to the incoming laser light (wavelength $\lambda$) modifies the incident angle $\theta$ and hence the periodicity can be tuned. The theoretical lower limit of 133 nm is defined by the wavelength used. The realistic upper limit of our setup is 1000 nm, since the area which is exposed with an interference pattern shrinks as $\theta$ approaches zero and the incoming light becomes increasingly parallel to the mirror. The substrates are aligned in a way that the line pattern is formed perpendicular to the gold contacts. During the development step the exposed photoresist is removed, leaving stripes of unexposed photoresist (Figure 1.iii). A change in periodicity causes a proportional change in the stripes width.

However, further control of their width can be achieved by adjusting the exposure energy dose. Since a positive photoresist is used, exposures with a higher energy dose by longer exposure times or increased laser power can be used to significantly narrow the stripes. Consequently, the width and density of nanochannels can be tuned independently within the mentioned periodicity limits (Figures 3a and b).

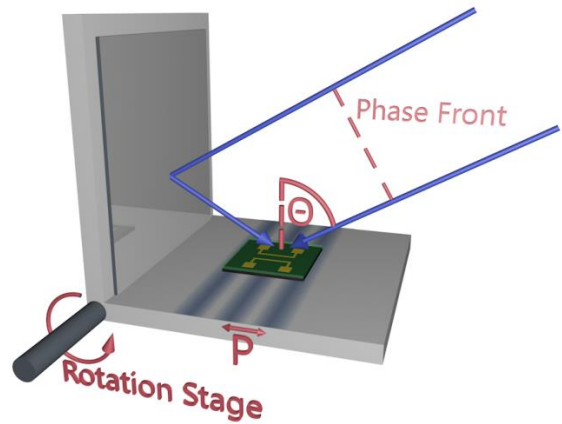

**Figure 2:** Laser interference lithography setup in the Lloyd's Mirror configuration. The blue arrows represent the incident light's k-vectors. The periodicity P is dependent on the angle $\theta$ between incident light and the sample's surface normal. It can be changed by rotating the stage with respect to the phase front.

Interference lithography is usually conducted in flat and polished substrates. Since our substrate is patterned with a thin gold electrode which can hamper the LIL process, we ensure the continuity of the photoresist stripes along the different parts of the electrodes by carefully tuning the multilayer stack. The sacrificial layer smoothes steps created by the gold electrodes and minimizes unwanted reflectivity changes near the electrodes. There are two additional reasons for incorporating this extra layer plus the reactive ion etching step (Figure 1.iv): RIE is used in order to achieve perfectly rectangular nanochannels and to improve the nanochannels' sidewall and upper surface quality, since photoresist patterns often exhibit a significant surface roughness. Most importantly, by etching under isotropic conditions the sacrificial polymer can be thinned in horizontal direction (Figure 3d). Using this *undercut* effect, feature sizes beyond the diffraction limit (Figure 3e) can be achieved.

However, the use of an additional sacrificial layer and a $SiO_2$ hard mask as described here is not mandatory and can be left out for easier synthesis. Güder *et al.* and Peeni *et al.* have presented similar approaches for nanochannel creation by directly using the patterned photoresist as sacrificial layer [36,42].

The pattern is transferred into the sacrificial layer by a sequential reactive ion etching procedure. The samples are first exposed to an oxygen plasma to remove the upper organic anti-reflective coating layer. The thereby exposed underlying $SiO_2$ hard mask is selectively etched by an $CHF_3$ plasma. Stripes of $SiO_2$ remain un-etched under the photoresist acting as a hard mask for a second oxygen plasma etching which transfers the line pattern into the sacrificial layer and removes the remaining photoresist on top of the silica hard mask. As a result, perfectly rectangular polymer stripes with a thin top layer of $SiO_2$ remain.

The patterned sacrificial layer is subsequently conformally coated by a thick layer of aluminum oxide deposited by ALD (Figure 1.v). ALD is ideally suited to mold the polymer stripes perfectly at low temperatures due to self-limiting surface reactions. The organic sacrificial layer is then completely removed from the channels by calcination in air (Figure 1.vi).

By a combination of overexposure during LIL and RIE under isotropic conditions, the smallest channels which were synthesized have a cross section of less than 40x40 nm² (Figure 3e), while still preserving the square shape and the periodicity with high homogeneity.

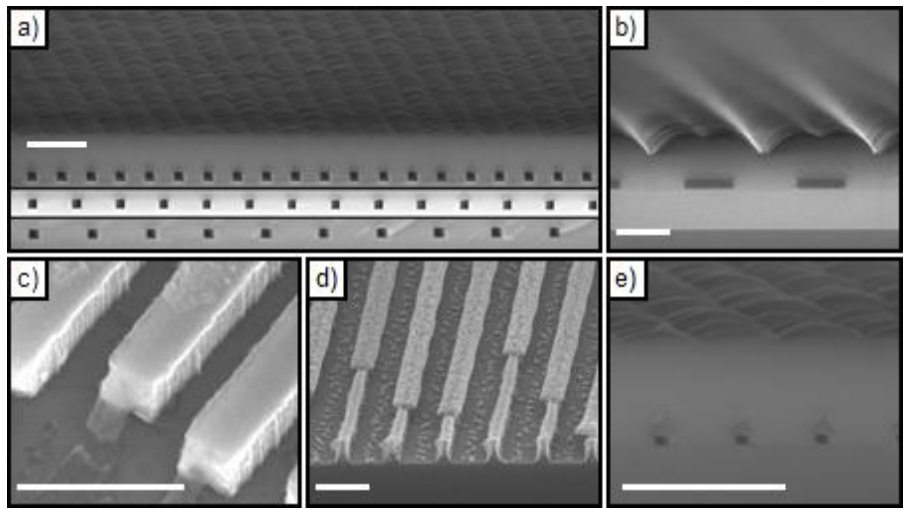

**Figure 3**: SEM images of square nanochannels and nanowires. All scale bars correspond to 400 nm. a) Overview of equally sized 80x80 nm² channels with inter-channel distances of 200 nm, 300 nm and 400 nm respectively, created by a combination of tilting the interferometer stage, overexposure and isotropic etching. b) Flat 80x350 nm² nanochannels with 850 nm periodicity c) Close up image of square 200x200 nm² nickel nanowires with 400 nm periodicity, d) polymer stripes which were thinned out horizontally due to RIE with higher chamber pressure, displaying an *undercut* under the $SiO_2$ hardmask and e) ultra-small nanochannels with dimensions less than 40x40 nm² and 200 nm periodicity.

In order to carry out the electrodeposition within the nanochannels, either simple mechanical scratching by hand and a scalpel or more accurate laser writing lithography followed by selective etching is used to open a small window in the desired place in the $Al_2O_3$ layer. The size of the window is tunable so that the amount of opened nanochannels (>5) can be defined at will. The window enables the electrolyte to enter the nanochannels (Figure 4a). Optical microscopy reveals that capillary force drags liquid along the nanochannels which are oriented perpendicular to the electrode front (Figure 5), here displayed on large nanochannels of approximately 250x250 nm² in cross section and 820 nm periodicity for easier optical imaging. We observe the same behavior in small channels; however the movement of the liquid is at much greater speed and cannot be captured.

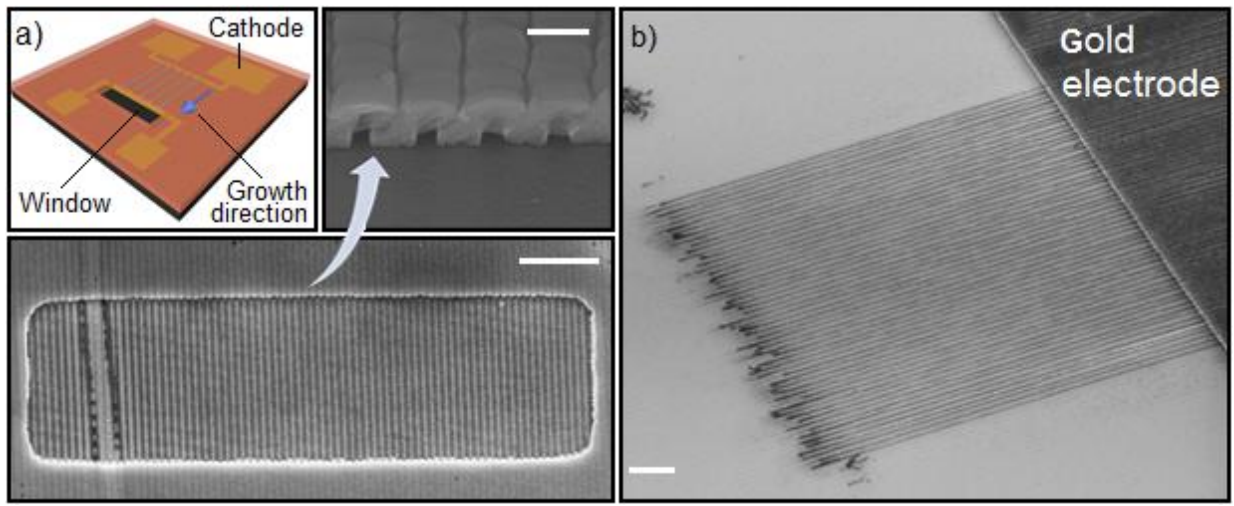

**Figure 4**: a) Schematic illustration and SEM images of an electrolyte window displaying selective opening of a limited number of nanochannels. Scale bars correspond to 400 nm (top) and 4 μm (bottom). The larger electrode in the schematic is used as a working electrode (cathode) in the electrodeposition process. b) Tilted SEM view (45°) of a parallel nanowire array. Scale bar is 2 μm.

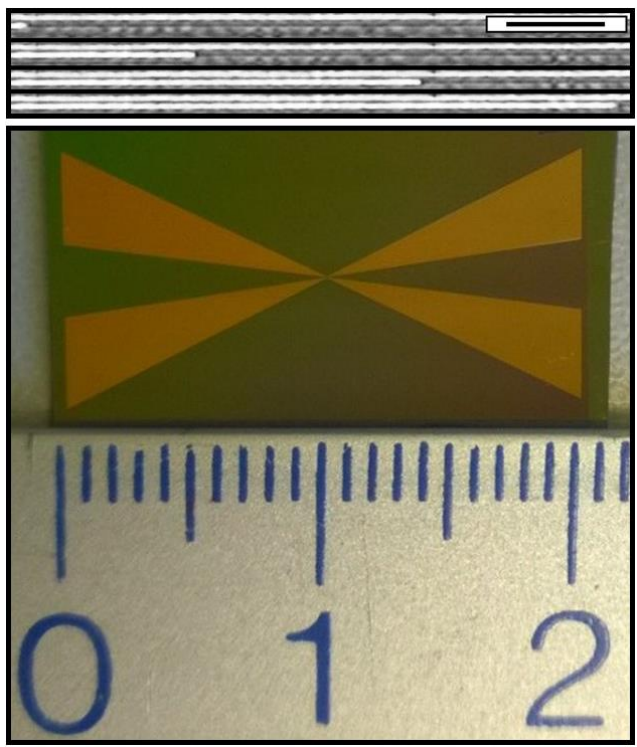

**Figure 5**: Top: Optical microscopy of liquid dragged along the nanochannels (250x250nm² with 800 nm periodicity) due to capillary force. Images are taken with 3 seconds in between each. Scale bar corresponds to 10 μm. Bottom: Digital photograph of a sample used for four-probe magnetoresistance measurements. The channels are oriented from top to bottom.

The gold electrode on the opposite side of the previously opened window is electrically connected as explained in the experimental section and used as the working electrode (Figure 4a) in a standard electrodeposition setup with three electrodes. Upon application of a correct voltage, the metal ions in the electrolyte are reduced at the working electrode and thus nanowire growth starts in direction towards the window. Pulsed electrodeposition is carried out to assure homogenous growth (Figure 4b) [19]. The deposition is time-controlled and continues until the nanowires reach the second electrode, assuring electrical contact for later measurements. The length of the wires is limited to approximately 100 μm, since the electrolyte circulation becomes more limited, the longer the distance is between the electrode and the window.

On the first sight our route might seem complex as compared to sputtering of the nanowires directly after the lithographic step [43]. There are several reasons why we believe our route offers great advantages: Achieving larger thicknesses in narrow pitches by sputtering is difficult, since it may prevent the lift-off procedure from working. Additionally, in periodic structures and as the feature sizes get smaller, the sputtered material tends to accumulate on top of the photoresist stripes rather than in the trenches [44]. By using electrodeposition, the wires can be precisely placed in the desired location. Most importantly, since the wires growth direction is along their long axis (as compared to growth along their short axis when sputtered), segmented nanowires with hundreds of repeats can be grown. These kinds of nanowires become especially important in spintronics, for example in integrated spin-valve, giant-magnetoresistance (GMR) and even tunnel-magnetoresistance (TMR) nanowires [23,24]. Various materials such as transition metals (Fe, Ni, Co, Cu) and semiconductors ($Bi_2Te_3$) have been grown inside the channels.

As a proof of our concept we present electrical measurements on iron nanowires (Figure 6). During the *in situ* measurement the nanowires remain within the protecting oxide template. The sample shows perfectly linear I-V relations at 2 K and 300 K demonstrating ohmic contacts between both gold electrodes and the iron nanowire (Figure 6a). From the perfectly linear $\rho(T)$ behavior between 50 K and 300 K indicated in the inset in Figure 6a, we infer that our nanowires are pure metallic. Since even short exposures of iron to ambient air lead to its oxidation, we prove that our template provides full protection against corrosion. If the template needs to be removed for SEM-imaging or sensing applications, weak 0.1 M KOH can be used.

Figure 6b shows the high-field temperature dependent magnetoresistance. The pronounced peak at 0 T is due to the anisotropic magnetoresistance effect, while the negative dependence between resistance and magnetic field for $B > 2$ T is attributed to a magnon freeze out [45,46]. An in-depth discussion of the temperature dependent magnon contribution and the electrical characteristics of electrodeposited Fe nanowires will follow in a future publication.

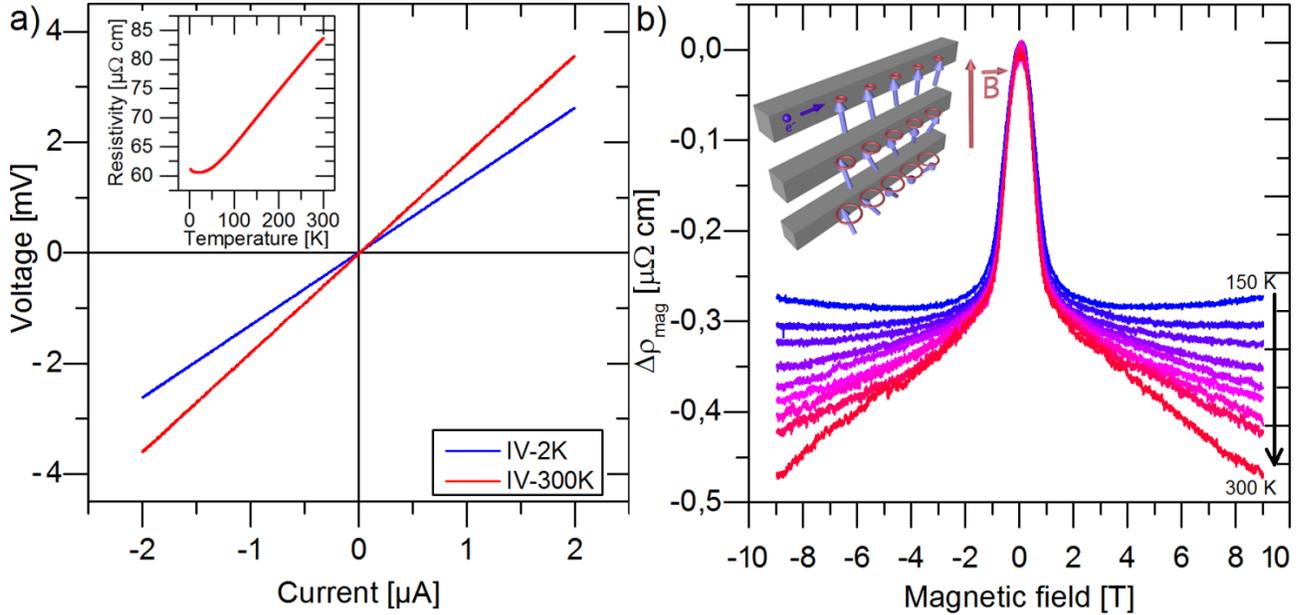

**Figure 6:** a) I-V curves of electrodeposited iron nanowires (70x80x12000 nm³) grown and characterized within the nanochannels template. The inset shows the temperature dependent resistivity which is approximately 9 times higher than the resistivity of bulk iron. This is a typical behavior for thin metallic nanowires due to boundary scattering. In publications on electrical properties of single cobalt and nickel nanowires, resistivities which are 7 times larger than bulk values were reported [47,48]. A small increase for T<15 K is observed which is attributed to localization effects. b) Temperature dependent magnetoresistance measurements from 150 K (blue) to 300 K (red). The cartoon displays the decrease of a spin-wave's amplitude from low magnetic fields (bottom) to high magnetic fields (top), resulting in reduced s-d scattering of the conduction electrons.

## 3. Conclusion

In summary, we have developed a technique for rapid synthesis of parallel electrodeposited nanowires by a combination of LIL, RIE and ALD. Our approach allows for the fabrication of highly resistant and versatile templates. With the proposed synthesis route, feature sizes well below the diffraction limit of the used wavelength can be achieved. In a model system we demonstrate the electrodeposition and electrical characterization of oxidation sensitive Fe nanowires. The presented approach can easily be extended to various application fields, such as biological and chemical sensors, racetrack memories, gate-voltage dependent measurement devices as well as electrical and magneto-optical characterization of 1D nanostructures. Wafer-scale arrays of well-positioned, parallel nanowires can be created, opening up the possibility for production of large amounts of nanowire-transistors and their incorporation into MOSFET designs.

While the fabrication of the template system alone is indeed more complex than producing anodic alumina templates or polycarbonate membranes, the discussed advantages in our opinion outweigh. In fact, the total amount of work necessary for fabrication, device creation and characterization of nanowires is strongly reduced. Additionally, the fabrication consists of only three major steps which are well established, fast, highly reproducible and compatible with "assembly-line" like production.

Finally, our approach can be used to create nanowires with tailor-made, rectangular magnetic and non-magnetic segments, thus selectively introducing stable transversal anisotropy within one nanowire. The possibility of using these as shift registers for the storage of multiple bits of data within one nanowire will be explored in a future publication.

## 4. Experimental Section

*4.1 Template fabrication.*

Si<100> wafers coated with 100 nm thick thermally grown silicon dioxide layer (*Si-Mat Inc.*) are used as substrates. For electrode patterning 500 nm *ma-P 1205* positive photoresist is used (*Microchemicals GmbH*) and illuminated with a direct lithography setup consisting of a 405 nm wavelength laser with 1μm spot and a movable stage (*μPG 101 LaserWriter* from *Heidelberg Instruments Mikrotechnik GmbH*). Electrodes are sputtered with 5 nm Cr as an adhesive promoter and 30 nm Au using a *308R Coating System (Cressington Scientific Instruments Ltd.)*. The layer stack in our experiments is built up by spin coating 35-90 nm *GenARC 266* (*Brewer Science Inc*.) or 120-250 nm *PMMA/MA 317.04* (*Allresist GmbH*) as a sacrificial layer, RF-Sputtering of 30 nm $SiO_2$ as a hard mask using a *CRC-300 (Torr International Inc.)*, spin coating 35-90 nm of *GenARC 266* as an antireflective coating for the LIL followed by 200 nm *UV2000* (*Rohm and Haas Electronic Materials Inc*.) positive photoresist. The required thickness of the antireflective coating is calculated using the program *Multilayer Reflectivity* by Michael E. Walsh in order to minimize the reflected power at the interface between photoresist and antireflective coating [39]. The exposure is conducted in a self-built LIL setup. The laser used in this work is a *FQCW 266-50 (CryLas GmbH)* with 266 nm wavelength and 50 mW output power. RIE is conducted with a *Si220* (*Sentech Instruments GmbH*) using an oxygen plasma for 60 seconds (10 mTorr chamber pressure, 10 sccm mass flow rate and 75 W power), followed by a $CHF_3$ plasma for 230 s (10 mTorr, 25 sccm $CHF_3$, 75 W) respectively. Depending on the thickness of the sacrificial layer, the last RIE step is conducted at the required times under above conditions. Increasing the chamber pressure during oxygen plasma etching to 25 mTorr as well as the etching time creates undercuts below the $SiO_2$ hardmask. ALD of $Al_2O_3$ is conducted in a *TFS200HP (BENEQ Oy)* reactor using tri-methyl aluminum and water precursors, both kept constant at 20°C with opening times of 250 ms and 500 ms respectively. The chamber temperature is set to 150 °C. The deposition is carried out in stop-mode operation with exposure times of 10 s, pump times of 10 s and approximately 1200 total cycles. A local electrolyte window either scratched mechanically by a scalpel or etched in 5% phosphoric acid at 45 °C for two hours before the sacrificial polymer is removed by calcination in air at 350 °C for at least 24 h at a heating rate of 1 °C/min.

*4.2 Nanowire growth.*

For the growth of iron nanowires by pulsed electrodeposition [19] an iron sulfate bath is used consisting of boric acid (4.5 g/l), glycine (12.2 g/l), ascorbic acid (1.0 g/l ) and iron(II) sulfate heptahydrate (4.5 g/l). Prior to electrodeposition, the electrolyte is bubbled with argon gas to avoid changes in the oxidation state of the iron ions [20]. A three-electrode cell with an Ag/AgCl/KCl$_{sat}$ reference electrode and a platinum wire as counter electrode is used. The electrode which is patterned on the substrate is connected on two ends and acts as the working electrode. One end serves as a conduction line for the current while the other serves as a measurement point of the potential directly at the location of nanowire growth with respect to the reference electrode. Thus, the resistance drop along the path of the thin electrodes which could cause a change in deposition potential is not taken into account. The pulse cycles consist of a deposition pulse at -1.3 V for 10 ms followed by an off-time of 100 ms at -0.45 V in which the ion concentration is allowed to regenerate, since there is no current flow. The bath is moderately stirred and kept at 20 °C. The horizontal growth rate under these conditions inside the channels is determined to be 300 nm/min. If the wires are to grow from one contact to another, electric contact on both ends is assured by controlling the deposition time with an ample buffer, since too long growth does not create any problems.

*4.3 Electric measurements.*

In order to connect the chip carrier with the sample a wire bonder is used. A thin aluminum wire is bonded to the electric contacts by applying force perpendicular to the surface plane combined with a short ultrasonic pulse. This proved to be sufficient to penetrate the layer of porous aluminum oxide channels which covers the contacts. If it is necessary to completely free the contact pads, for example if probes are used for measurement, the edges of the sample can be dipped into KOH (25 %$_{wt.}$) for 3 min. It is important to make sure the channels are aligned parallel to the etching liquid surface, to make sure capillary forces do not drag the solution into the matrix, possibly reaching the grown wires. If the contact structure is made large enough in beforehand (approx. 2x2 cm²), this procedure can be done with ease and without risk of damaging the wires (example structure see Figure 5). In our example we use a four-point measurement setup to characterize the magnetoresistance of a single Fe-nanowire using a *PPMS Dynacool* (*Quantum Design Inc.*) cryostat. SEM characterization (*Zeiss Sigma*) reveals that only one nanowire is connecting both sides of the electrodes, since the growth rates in neighboring pores was slightly slower. The surrounding template does not need to be removed for imaging, since the contrast difference between a pore that contains a wire and an empty one is sufficient. For the measurements, an AC current (lock in freq. 128.174 Hz) is injected while the voltage drop over the nanowire is measured. The magnetic field is applied perpendicular to the nanowire.

**Funding Sources**

The research leading to these results has received funding from the European Unions's 7th Framework Programme under grant agreement n°309589 (M3d) and we gratefully acknowledge financial support from the German Research Foundation (DFG) via SFB 986 "M3", project C3.

**Acknowledgements**

We would like to thank Johannes Gooth and Heiko Reith for thoughtful discussions and assistance with electrical measurements.

**References**

[1]  Cui Y and Lieber C M 2001 *Science* **291** 851–3

[2]  Cui Y, Zhong Z H, Wang D L, Wang W U and Lieber C M 2003 *Nano Lett.* **3** 149–52

[3]  Duan X F, Huang Y, Cui Y, Wang J F and Lieber C M 2001 *Nature* **409** 66–9

[4]    Lavrijsen R, Lee J H, Fernandez-Pacheco A, Petit D C M C, Mansell R and Cowburn R P 2013 *Nature* **493** 647–50

[5]    Allwood D A, Xiong G, Faulkner C C, Atkinson D, Petit D and Cowburn R P 2005 *Science* **309** 1688–92

[6]    Cowburn R P 2009 *Nanoscale Magnetic Materials and Applications,* Eds. Ping Liu J, Fullerton E, Gutfleisch O and Sellmyer D J, New York, Springer, 219–36

[7]    Hicks L D and Dresselhaus M S 1993 *Phys. Rev. B* **47** 16631–4

[8]    Wu P M, Gooth J, Zianni X, Svensson S F, Gluschke J G, Dick K A, Thelander C, Nielsch K and Linke H 2013 *Nano Lett.* **13** 4080–6

[9]    Kwiat M, Cohen S, Pevzner A and Patolsky F 2013 *Nano Today* **8** 677–94

[10]    Tsivion D, Schvartzman M, Popovitz-Biro R, von Huth P and Joselevich E 2011 *Science* **333** 1003–7

[11]    Menke E J, Thompson M A, Xiang C, Yang L C and Penner R M 2006 *Nat. Mater* **5** 914–9

[12]    Moreno i Codinachs L, Birkenstock C, Garma T, Zierold R, Bachmann J, Nielsch K, Schoening M J and Fontcuberta i Morral A 2009 *Phys. Status Solidi A* **206** 435–1

[13]    Xiang Y, Keilbach A, Codinachs L M, Nielsch K, Abstreiter G, Fontcuberta i Morral A and Bein T 2010 *Nano Lett.* **10** 1341–46

[14]    Yang F, Taggart D K and Penner R M 2009 *Nano Lett.* **9** 2177–82

[15]    Barth S, Hernandez-Ramirez F, Holmes J D and Romano-Rodriguez A 2010 *Prog. Mater. Sci.* **55** 563–627

[16]    Duan X F, Wang J F and Lieber C M 2000 *Appl. Phys. Lett.* **76** 1116–8

[17]   Hamdou, B, Kimling J, Dorn A, Pippel E, Rostek R, Woias P and Nielsch K 2013 *Adv. Mater.* **25** 239–44

[18]   Schio P, Vidal F, Zheng Y, Milano J, Fonda E, Demaille D, Vodungbo B, Varalda J, de Oliveira A J A and Etgens V H 2010 *Phys. Rev. B* **82** 094436

[19]   Nielsch K, Müller F, Li A P and Gösele U 2000 *Adv. Mater.* **12** 582–6

[20]   Salem M S, Sergelius P, Zierold R, Moreno J M M, Görlitz D and Nielsch K 2012 *J. Mater. Chem.* **22** 8549–57

[21]   Salem M S, Sergelius P, Corona R M, Escrig J, Görlitz D and Nielsch K 2013 *Nanoscale* **5** 3941–47

[22]   Sander M S, Prieto A L, Gronsky R, Sands T and Stacy A M 2002 *Adv. Mater.* **14** 665–7

[23]   Blondel A, Meier J P, Doudin B and Ansermet J P 1994 *Appl. Phys. Lett.* **65** 3019–21

[24]   Maqableh M M, Huang X, Sung S-Y, Reddy K S M, Norby G, Victora R H and Stadler B J H 2012 *Nano Lett.* **12** 4102–9

[25]   Voelklein F, Reith H and Meier A 2013 *Phys. Status Solidi A* **210** 106–18

[26]   Bässler S, Böhnert T, Gooth J, Schumacher C, Pippel E and Nielsch K 2013 *Nanotechnology* **24** 495402

[27]   Weber D P, Ruffer D, Buchter A, Xue F, Russo-Averchi E, Huber R, Berberich P, Arbiol J, Morral A F I, Grundler D and Poggio M 2012 *Nano Lett.* **12** 6139–44

[28]   Mavrokefalos A, Moore A L, Pettes M T, Shi L, Wang W and Li X 2009 *J. Appl. Phys.* **105** 104318

[29]   Cronin S B, Lin Y M, Rabin O, Black M R, Dresselhaus G, Dresselhaus M S and Gai P L 2002 *Microsc. Microanal.* **8** 58–63

[30]   Heiderich S, Toellner W, Boehnert T, Gluschke J G, Zastrow S, Schumacher C, Pippel E and Nielsch K 2013 *Phys. Status Solidi RRL* **7** 898–902

[31]   Menard L D and Ramsey J M 2011 *Nano Lett.* **11** 512–7

[32]   Whitesides G M 2006 *Nature* **442** 368–73

[33]   Mijatovic D, Eijkel J C T and van den Berg A 2005 *Lab Chip* **5** 492–500

[34]  Guan W, Li S X and Reed M A 2014 *Nanotechnology* **25** 122001

[35]   Pevzner A, Engel Y, Elnathan R, Tsukernik A, Barkay Z and Patolsky F 2012 *Nano Lett.* **12** 7–12

[36]   Gueder F, Yang Y, Krueger M, Stevens G B and Zacharias M 2010 *ACS Appl. Mater. Interfaces* **2** 3473–8

[37]   Li W L, Tegenfeldt J O, Chen L, Austin R H, Chou S Y, Kohl P A, Krotine J and Sturm J C 2003 *Nanotechnology* **14** 578–83

[38]   Tas N R, Berenschot J W, Mela P, Jansen H V, Elwenspoek M and van den Berg A 2002 *Nano Lett.* **2** 1031–32

[39]   Walsh M E 2000 *Nanostructuring Magnetic Thin Films Using Interference Lithography*; Master's Thesis, Massachusets Institute of Technology

[40]   Montero Moreno J M, Waleczek M, Martens S, Zierold R, Görlitz D, Vega Martinez V, Prida V M and Nielsch K 2013 *Adv. Funct. Mater.* **24** 1857-63

[41]   de Boor J, Kim D S and Schmidt V 2010 *Opt. Lett.* **35** 3450–2

[42]   Peeni B A, Conkey D B, Barber J P, Kelly R T, Lee M L, Woolley A T and Hawkins A R 2005 *Lab Chip* **5** 501–5

[43]   Park J-M, Nalwa K S, Leung W, Constant K, Chaudhary S and Ho K-M 2010 *Nanotechnology* **21** 215301

[44]   Yeon J, Lee Y J, Yoo D E, Yoo K J, Kim J S, Lee J, Lee J O, Choi S-J, Yoon G-W, Lee D W, Lee G S, Hwang H C and Yoon J-B 2013 *Nano Lett.* **13** 3978–84

[45]   Mott N F 1964 *Adv. Phys.* **13** 325

[46]   Raquet B, Viret M, Sondergard E, Cespedes O and Mamy R 2002 *Phys. Rev. B* **66** 024433

[47]   Fernandez-Pacheco A, De Teresa J M, Cordoba R and Ibarra M R 2009 *J. Phys. D: Appl. Phys.* **42** 055005

[48]   Ou M N, Yang T J, Harutyunyan S R, Chen Y Y, Chen C D and Lai S J 2008 *Appl. Phys. Lett.* **92** 063101

Graphical Abstract

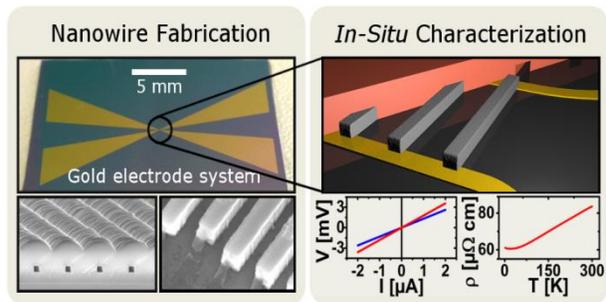

An array of parallel nanochannels is used as a template for electrochemical synthesis of nanowires. This newly developed integrated device allows for a direct electrical characterization of the nanowires in the place of their growth without any further processing steps. The nanochannels have a square cross section and dimensions of as low as 40x40 nm